\documentclass[12pt]{article}

\newcommand{\tr}{{\rm tr}}

\begin{document}
\vspace*{-.6in}
\thispagestyle{empty}
\begin{flushright}
DAMTP R96/49\\
CALT-68-2077\\
hep-th/9611065
\end{flushright}
\baselineskip = 20pt

\vspace{.5in}
{\Large
\begin{center}
Interacting Chiral Gauge Fields in Six Dimensions\\ and Born--Infeld Theory
\end{center}}
\vspace{.4in}

\begin{center}
Malcolm Perry\\
\emph{DAMTP, University of Cambridge, Cambridge, CB3 9EW, U.K.}
\end{center}

\begin{center}
and
\end{center}

\begin{center}
John H. Schwarz\footnote{Work supported in part by the U.S. Dept. of Energy 
under Grant No. DE-FG03-92-ER40701.} \\
\emph{California Institute of Technology, Pasadena, CA  91125 USA}
\end{center}
\vspace{1in}

\begin{center}
\textbf{Abstract}
\end{center}
\begin{quotation}
\noindent Dimensional reduction of a self-dual tensor gauge field in 6d gives
an Abelian vector gauge field in 5d.  We derive the conditions under which 
an interacting 5d theory of an Abelian vector gauge field is the dimensional
reduction of  a 6d Lorentz invariant interacting theory of a self-dual tensor.  
Then we specialize to the particular 6d theory that gives 5d Born--Infeld theory.
The field equation and Lagrangian of this 6d theory are formulated with manifest 
5d Lorentz invariance, while the remaining Lorentz symmetries are realized 
nontrivially.  A string soliton with finite tension and self-dual charge is constructed.
\end{quotation}
\vfil

\newpage

\pagenumbering{arabic} 

\section{Introduction}

There are three classes of super $p$-branes that occur in string theory and M
theory.  The first class of $p$-branes, which includes superstrings and the M
theory supermembrane, have world-volume theories whose physical degrees of
freedom consist only of scalars and spinors.  These theories were classified in
the original ``brane scans,'' \cite{achucarro}
and their world-volume actions were constructed
some time ago.\cite{duff}  More recently, attention has focussed on $D$-branes, which have
also been classified.\cite{polchinski}  A characteristic feature of their world volume theories
is the presence of a $U(1)$ gauge field whose self interactions are given by
Born--Infeld theory.\cite{born,fradkin,abouelsaoud,bergshoeff}  
The third class of $p$-branes, exemplified by the M theory
five-brane, has a second-rank tensor gauge field in the world volume theory.\cite{callan}

The physical degrees of freedom of the 6d world volume theory of the M theory
five-brane consist of a $N = (2,0)$ tensor supermultiplet.  This multiplet
contains a two-form $B_{MN}$, with a self-dual field strength, five scalars,
and two chiral spinors.  The scalars and spinors can be interpreted as
Goldstone bosons and fermions associated with broken translation symmetries and
supersymmetries.  When 11d M theory is compactified on a circle it gives 10d
type IIA superstring theory.  Some of the $p$-branes of the IIA theory have a
simple M theory interpretation.  In particular, wrapping one dimension of the M
theory five-brane on the compact spatial dimension gives the four-brane of IIA
theory.  This four-brane is a $D$-brane and therefore its world volume theory
consists of a $U(1)$ gauge field plus scalars and spinors, and the $U(1)$ gauge
field has Born--Infeld self interactions.  This 5d world volume
theory must arise as the dimensional reduction of the 6d
five-brane world-volume theory.  Thus, the five-brane world volume theory must
be a self-interacting theory of the $N=(2,0)$ tensor supermultiplet.  Our goal
is to construct this theory.\footnote{6d theories in which tensor
supermultiplets interact with other matter supermultiplets have been considered
in Ref. \cite{bergshoeff2}. As far as we know, theories of self-interacting self-dual
tensors have not been proposed previously.}

In this paper, as a first step towards understanding the M theory five-brane, 
we simplify the problem by dropping all
scalars and spinors, thereby giving up supersymmetry.  So our problem is to
construct a 6d Lorentz invariant interacting theory of a self-dual
tensor gauge field that gives Born--Infeld theory upon reduction to 5d.  
Actually, we will do something a bit more general.  We will only
assume that the 5d theory has an action that is an arbitrary
Lorentz invariant function of $F_{\mu\nu}$, and reduces to Maxwell theory
$(F^2)$ for weak fields.  Then we will examine the conditions for ``lifting''
this to a 6d Lorentz invariant theory of a chiral tensor field.
Although there is a large class of interacting Lorentz invariant 6d theories
that can be constructed in this way, the one that reduces to Born-Infeld in 5d
is particularly simple. This is fortunate, since it  is also the one we are 
most interested in. 

One well-known issue that makes the analysis challenging is the lack of a
manifestly covariant action for theories with chiral bosons.\cite{marcus}
\footnote{This statement assumes a formulation of the theory with a
finite number of fields. By adding an infinite number of auxiliary fields it is
apparently possible to circumvent this conclusion. For a recent discussion of
such an approach see \cite{berk}.}  
In the case of
type IIB supergravity, for example, there is no action with manifest
10d general covariance, though covariant field equations do exist.\cite{schwarz}
The theory we are seeking here is simpler than type IIB supergravity in as much
as it is just a flat-space matter theory.  However, it has a surprising new
feature.  It appears that not only is there no manifestly Lorentz invariant
action, but {\it even the field equation lacks manifest Lorentz invariance}. 
This may sound
rather disturbing, but it is not really so bad.  We are able to exhibit field
equations and an action with manifest 5d Lorentz invariance and
to prove invariance under Lorentz transformations mixing those five dimensions
with the sixth one.  This theory is formulated entirely in terms of a gauge field
$B_{\mu\nu}$, where $\mu, \nu$ are 5d indices.

Theories with a two-form gauge field $B_{\mu\nu}$ are natural candidates for
having string-like solitons (one-branes).  For example, in 10d cases, not only are
supergravity theories the low-energy effective descriptions of the corresponding string
theories, but the strings themselves can be reconstructed, at least
approximately, as classical soliton solutions of the supergravity field
equations.  In the case of 6d, there is a great deal of evidence for a new class of string
theories -- non-critical self-dual strings.\cite{witten}  These are
non-gravitational theories defined in six flat dimensions.  Moreover, the
massless spectrum of such strings always contains a chiral two-form
gauge field.  Thus it is natural to examine our field equations for a
string-like soliton.  We find that there is one with the expected properties: its tension is finite,
and it carries a self-dual charge.   In the case of 10d we usually regard the
superstring as fundamental and supergravity as derived.  In the 6d
case, it may make more sense to consider the  field theory as fundamental
and the string as derived.

In Section 2 we describe the free theory in considerable detail.  In this
setting all the subtle issues relating to Lorentz invariance already appear.
Section 3 then formulates the interacting field equations and derives the conditions for
Lorentz invariance in 6d. The particular example that
reduces to Born--Infeld theory in 5d is identified and described. The subsequent 
analysis is restricted to that example, since it is the one that is relevant for
the M theory five-brane application that we have in mind.
Section 4 presents the 6d action with manifest
5d Lorentz invariance and proves that it has the symmetry
required for complete 6d Lorentz invariance.  Section 5 presents
the string soliton.  Section 6 summarizes our conclusions and suggests directions for future research.

\section{The Free Theory}

In this section we describe a free self-dual tensor gauge field in 6d 
and the free Maxwell theory in 5d that is obtained by
dimensional reduction.  We denote 5d coordinates by $x^\mu =
(x^0, x^1, \ldots , x^4)$ and 6d ones by $x^M = (x^\mu, x^5)$.
The Lorentz metrics in 5d and 6d are $\eta^{\mu\nu} = (- ++++)$
and $\eta^{MN} = (-+++++)$.  The invariant antisymmetric tensors $\epsilon^{\mu_{1}
\ldots \mu_{5}}$ and $\epsilon^{M_{1} \ldots M_{6}}$ have $\epsilon^{01234} = -
\epsilon_{01234} = 1$ and $\epsilon^{012345} = - \epsilon_{012345} = 1$.

The 6d gauge field $B_{MN}$ has a three-form field strength
\begin{equation}
H_{MNP} = \partial_M B_{NP} + \partial_N B_{PM} + \partial_P B_{MN},
\end{equation}
which is invariant under the usual gauge transformations $(\delta B_{MN} =
\partial_M \lambda_N - \partial_N \lambda_M)$.  The dual field strength is
defined to be
\begin{equation}
\tilde{H}^{MNP} = {1\over 6} \epsilon^{MNPQRS} H_{QRS}.
\end{equation}
The self-duality condition
\begin{equation}
\tilde{H}_{MNP} = H_{MNP} \label{selfdual}
\end{equation}
is a first order field equation for a free chiral boson. 
The Lorentzian signature of the 6d spacetime guarantees that
the field $H_{MNP}$ is real.
The field equations have manifest 6d Lorentz
invariance, a feature that will be sacrificed when interactions are included.

Let us decompose the above into 5d pieces.  $B_{MN}$ gives rise to $B_{\mu\nu}$
 and $A_\mu \equiv B_{\mu 5}$.  We define $F_{\mu\nu} = \partial_\mu A_\nu -
\partial_\nu A_\mu$, as usual.  Then $H_{MNP}$ decomposes into
\begin{equation}
H_{\mu\nu\rho} = \partial_\mu B_{\nu\rho} + \partial_\nu B_{\rho\mu} +
\partial_\rho B_{\mu\nu}
\end{equation}
and
\begin{equation}
{\cal F}_{\mu\nu} \equiv H_{\mu\nu 5} = F_{\mu\nu} + \partial_5 B_{\mu\nu}.
\end{equation}
We also define
\begin{equation}
\tilde{H}^{\mu\nu} = {1\over 6} \epsilon^{\mu\nu\rho\lambda\sigma}
H_{\rho\lambda\sigma},
\end{equation}
whose inversion is
\begin{equation}
H^{\mu\nu\rho} = - {1\over 2} \epsilon^{\mu\nu\rho\lambda\sigma}
\tilde{H}_{\lambda\sigma}.
\end{equation}
The minus sign is a consequence of Lorentzian signature.  The self-duality
equation (\ref{selfdual}) in this notation becomes
\begin{equation}
\tilde{H}_{\mu\nu} = {\cal F}_{\mu\nu}. \label{fivedeqn}
\end{equation}

Since $\tilde{H}_{\mu\nu} = {\cal F}_{\mu\nu}$ is just a rewriting of
$\tilde{H}_{MNP} = H_{MNP}$, we already know that it has 6d Lorentz invariance.
However, to set the stage for the next section, it is useful to prove this
directly.  Since 5d covariance is manifest, we only examine transformations
mixing the $\mu$ directions with the 5 direction, calling the infinitesimal
 parameters
$\Lambda_\mu$.  As usual, a Lorentz transformation has an \lq\lq orbital''
 part and a
\lq\lq spin''  part.  The orbital part is given by the operator
\begin{equation}
\Lambda \cdot L = (\Lambda \cdot x) \partial_5 - x_5 (\Lambda \cdot \partial).
\end{equation}
Decomposing the standard 6d Lorentz transformation formulae into 5d pieces
 one has
\begin{equation}
\delta B_{\mu\nu} = (\Lambda \cdot L) B_{\mu\nu} + \Lambda_\nu A_\mu -
\Lambda_\mu A_\nu, \label{btrans}
\end{equation}
which implies
\begin{equation}
\delta H_{\mu\nu\rho} = (\Lambda \cdot L) H_{\mu\nu\rho} + \Lambda_\mu {\cal
F}_{\nu\rho} + \Lambda_\nu {\cal F}_{\rho\mu} + \Lambda_\rho {\cal F}_{\mu\nu}
\end{equation}
or, equivalently
\begin{equation}
\delta \tilde{H}^{\mu\nu} = (\Lambda \cdot L) \tilde{H}^{\mu\nu} + {1\over 2}
\epsilon^{\mu\nu\rho\lambda\sigma} \Lambda_\rho {\cal F}_{\lambda\sigma}.
\end{equation}
One also has
\begin{equation}
\delta A_\mu = (\Lambda \cdot L) A_\mu - \Lambda^\nu B_{\mu\nu},
\end{equation}
which implies
\begin{equation}
\delta {\cal F}_{\mu\nu} = (\Lambda \cdot L) {\cal F}_{\mu\nu} - \Lambda^\rho
H_{\mu\nu\rho}.
\end{equation}

We can now examine the effect of applying a Lorentz transformation to the
equation $\tilde{H}_{\mu\nu} - {\cal F}_{\mu\nu} = 0$.  The requirement of
invariance is that the variation should vanish using this equation.  In fact,
this works separately for the orbital and spin parts of the Lorentz
transformation
\[
\delta_{orb} (\tilde{H}_{\mu\nu} - {\cal F}_{\mu\nu}) = \Lambda \cdot L
(\tilde{H}_{\mu\nu} - {\cal F}_{\mu\nu}) = 0\]
\begin{equation}
\delta_{spin} (\tilde{H}_{\mu\nu} - {\cal F}_{\mu\nu}) = -{1\over 2}
\epsilon_{\mu\nu\rho\lambda\sigma} \Lambda^\rho (\tilde{H}^{\lambda\sigma} -
{\cal F}^{\lambda\sigma}) = 0.
\end{equation}
In the interacting theory the orbital symmetry will again work trivially,
but the spin part will require a careful analysis.

Let us now consider dimensional reduction to 5d.  This entails
setting $\partial_5 B_{\mu\nu} = 0$ in the above, so that the field equation
(\ref{fivedeqn}) becomes $\tilde{H}_{\mu\nu} = F_{\mu\nu}$.  Now $\partial^\mu
\tilde{H}_{\mu\nu} = 0$ is a Bianchi identity, so we obtain $\partial^\mu
F_{\mu\nu} = 0$ as a second order field equation involving only the field
$A_\mu$.  As expected, this is just Maxwell theory, which follows from a 5d
Lagrangian ${\cal L}_5 \sim F^{\mu\nu} F_{\mu\nu}$.

In 6d we can also convert to a second-order field equation by
utilizing a Bianchi identity.  We have $F_{\mu\nu}  = \tilde{H}_{\mu\nu} -
\partial_5 B_{\mu\nu}$, and thus we obtain
\begin{equation}
\epsilon^{\mu\nu\rho\lambda\sigma} \partial_\rho (\tilde{H}_{\lambda\sigma} -
\partial_5 B_{\lambda\sigma}) = 0.
\end{equation}
This field equation follows from the 6d action
\begin{equation}
S_6 = {1 \over 2}
 \int ( \tilde{H}^{\mu\nu} \partial_5 B_{\mu\nu} - \tilde{H}^{\mu\nu}
\tilde{H}_{\mu\nu}) d^6 x. \label{freeaction}
\end{equation}
Note that $S_6 $ is gauge invariant up to the integral of a total
 derivative, which
is good enough for suitable boundary conditions. This action is of
 the type introduced in
ref. \cite{henneaux}, which has also been discussed in 
refs. \cite{schwarz2,verlinde}.

We already know that $S_6 $ gives field equations with 6d Lorentz invariance.
Still, it is interesting to examine its symmetry directly.  Since the field
$A_\mu$ does not appear in $S_6$, it is convenient, but not essential, to
utilize the $A_\mu = 0$ gauge.  In this gauge the Lorentz
transformation in eq. (\ref{btrans}) simplifies to
\begin{equation}
\delta B_{\mu\nu} = (\Lambda \cdot L) B_{\mu\nu} = (\Lambda \cdot x) \partial_5
B_{\mu\nu} - x_5 (\Lambda \cdot \partial) B_{\mu\nu} , \label{btrans1}
\end{equation}
which is a symmetry of the $A_\mu = 0$ gauge field equation $\partial_5
B_{\mu\nu} = \tilde{H}_{\mu\nu}$.  While this equation is invariant under the
transformation (\ref{btrans1}), the action $S_6$ is not.  To get the right expression, we must
use the field equation to modify the transformation law as follows:
\begin{equation}
\delta B_{\mu\nu} = (\Lambda \cdot x) \tilde{H}_{\mu\nu} - x_5 (\Lambda \cdot
\partial) B_{\mu\nu}.
\end{equation}
The claim is that this describes a symmetry of $S_6$.

It is instructive to examine this claim explicitly:
\[
\delta S_6 \sim \int d^6 x \delta B_{\mu\nu} \epsilon^{\mu\nu\rho\lambda\sigma}
\partial_\rho (\tilde{H}_{\lambda\sigma} - \partial_5 B_{\lambda\sigma})\]
\begin{equation}
= \int d^6 x \epsilon^{\mu\nu\rho\lambda\sigma} ((\Lambda \cdot x)
\tilde{H}_{\mu\nu} - x_5 (\Lambda \cdot \partial) B_{\mu\nu}) \partial_\rho
(\tilde{H}_{\lambda\sigma} - \partial_5 B_{\lambda \sigma}).
\end{equation}
Multiplying this out, there are four terms, which we examine separately:
\begin{eqnarray}
\epsilon^{\mu\nu\rho\lambda\sigma} (\Lambda \cdot x) \tilde{H}_{\mu\nu}
\partial_\rho \tilde{H}_{\lambda\sigma}
&=& - {1\over 2} \epsilon^{\mu\nu\rho\lambda\sigma} \Lambda_\rho
\tilde{H}_{\mu\nu} \tilde{H}_{\lambda\sigma} + {\rm tot. \, deriv. }
 \nonumber \\
&=& \Lambda_\rho \tilde{H}_{\mu\nu} H^{\mu\nu\rho} + 
{\rm tot. \, deriv. } \nonumber \\
&=& \tilde{H}_{\mu\nu} (\Lambda \cdot \partial) B^{\mu\nu} +
 {\rm tot. \, deriv. }\\
- \epsilon^{\mu\nu\rho\lambda\sigma} (\Lambda \cdot x) \tilde{H}_{\mu\nu}
\partial_\rho \partial_5 B_{\lambda\sigma}
&=& - {1\over 2} (\Lambda \cdot x) \tilde{H}_{\mu\nu} \partial_5
\tilde{H}^{\mu\nu} = {\rm tot. \, deriv. } \\
- \epsilon^{\mu\nu\rho\lambda\sigma} x_5 (\Lambda \cdot \partial) B_{\mu\nu}
\partial_\rho \tilde{H}_{\lambda\sigma}
&=& 2 x_5 (\Lambda \cdot \partial) B_{\mu\nu} \partial_\rho
H^{\mu\nu\rho}\nonumber \\
&=& 2 x_5 (\Lambda \cdot \partial) B_{\mu\nu} 
((\partial \cdot \partial) B^{\mu\nu}
+ 2 \partial_\rho \partial^\mu B^{\nu\rho})\nonumber \\
&=& {\rm tot. \, deriv. }\\
\epsilon^{\mu\nu\rho\lambda\sigma} x_5 (\Lambda \cdot \partial) B_{\mu\nu}
\partial_\rho \partial_5 B_{\lambda\sigma}&=&
 - {1\over 2} \epsilon^{\mu\nu\rho\lambda\sigma} (\Lambda \cdot \partial)
B_{\mu\nu} \partial_\rho B_{\lambda\sigma} + {\rm tot. \, deriv. }\nonumber \\
&=& - \tilde{H}_{\mu\nu} (\Lambda \cdot \partial) B^{\mu\nu} +
 {\rm tot. \, deriv. }
\end{eqnarray}
Thus, up to total derivatives, two of the terms vanish and the other two
cancel.

If one computes the algebra $[\delta(\Lambda_1), \delta (\Lambda_2)]
B_{\mu\nu}$ the result consists of the expected 5d Lorentz transformation
plus a gauge transformation plus terms that vanish using the equations of
motion.  This is exactly the situation that is familiar in the case of
supersymmetric theories with incomplete off-shell supermultiplets.  There
seems to be no fundamental reason to demand better for the Lorentz group.

\section{The Interacting Theory}

Now let us examine the possibilities for extending the free theory of Section 2
to an interacting theory.  We take as our starting point the 5d $U(1)$ gauge
theory that arises upon dimensional reduction.  We assume that the 5d
Lagrangian is a function of the field strengths, but not their derivatives.
Then, since the 5d Lorentz group has rank two, Lorentz invariance implies that
the Lagrangian ${\cal L}_5 \sim f (y_1, y_2)$, where
\begin{eqnarray}
y_1 &\equiv& {1\over 2} \tr F^2 =
 - {1\over 2} F_{\mu\nu} F^{\mu\nu}\nonumber \\
y_2 &\equiv& {1\over 4} \tr F^4.
\end{eqnarray}
The classical field equation is $\partial_\mu \left({\delta S\over \delta
F_{\mu\nu}}\right) = 0$, which we ``solve'' by setting
\begin{equation}
\tilde{H}_{\mu\nu} = {\delta S\over\delta F_{\mu\nu}} = F_{\mu\nu} f_1 +
(F^3)_{\mu\nu} f_2,
\end{equation}
where $f_i \equiv {\partial f\over \partial y_i}$.  We match onto the free
theory by requiring that $f$ is analytic at $y_1 = y_2 = 0$ and
\begin{equation}
f  (y_1, y_2) = y_1 + O (y_1^2, y_2).
\end{equation}

Now we want a 6d theory that agrees with this upon dimensional reduction.
Since dimensional reduction eliminates $\partial_5$ terms, we must guess how to
add them in.  Fortunately, in this case, there is only one plausible guess that
is dictated by gauge invariance.  Namely, $F_{\mu\nu} \rightarrow {\cal
F}_{\mu\nu} = F_{\mu\nu} + \partial_5 B_{\mu\nu}$. 
 So we conjecture the 6d field
equation
\begin{equation}
\tilde{H}_{\mu\nu} = {\cal F}_{\mu\nu} f_1 + ({\cal F}^3)_{\mu\nu} f_2 ,
 \label{fulleqn} \end{equation}
where it is now understood that $y_1 = {1\over 2} \tr {\cal F}^2$ and $y_2 =
{1\over 4} \tr {\cal F}^4$.

The next step is to examine the transformation of the field equation under a
Lorentz transformation
\begin{eqnarray}
\delta \tilde{H}^{\mu\nu} &=& (\Lambda \cdot L) \tilde{H}^{\mu\nu} + {1\over 2}
\epsilon^{\mu\nu\rho\lambda\sigma} \Lambda_\rho {\cal
F}_{\lambda\sigma}\nonumber \\
\delta {\cal F}_{\mu\nu} &=& (\Lambda \cdot L) {\cal F}_{\mu\nu} - \Lambda^\rho
H_{\mu\nu\rho},
\end{eqnarray}
the same formulae as in the free theory.  Since $f$ only depends on ${\cal
F}_{\mu\nu}$ and not its derivatives, the orbital part of the Lorentz
transformation just gives $\Lambda \cdot L$ acting on the equation, thus
leaving the equation invariant.  Therefore, we need only examine the spin
parts.  The varied equation is
\begin{eqnarray}
{1\over 2} \epsilon^{\mu\nu\rho\lambda\sigma}
 \Lambda_\rho {\cal F}_{\lambda\sigma} =
&-& H^{\mu\nu\rho} \Lambda_\rho f_1 - \Lambda_\rho H^{\mu\alpha\rho} 
{\cal F}_{\alpha\beta} {\cal F}^{\beta\nu} f_2\nonumber \\
&-& \Lambda^\rho {\cal F}^{\mu\alpha} H_{\alpha\beta\rho} {\cal F}^{\rho\nu}
f_2 - \Lambda_\rho {\cal F}^{\mu\alpha} {\cal F}_{\alpha\beta} H^{\beta\nu\rho}
f_2\nonumber \\
&+& {\cal F}^{\mu\nu} \Lambda^{\rho} H_{\alpha\beta\rho} {\cal F}^{\alpha\beta}
f_{11} + ({\cal F}^3)^{\mu\nu} \Lambda^\rho H_{\alpha\beta\rho} {\cal
F}^{\alpha\beta} f_{12} \nonumber \\
&+& {\cal F}^{\mu\nu} \Lambda^\rho H_{\alpha\beta\rho} ({\cal
F}^3)^{\alpha\beta} f_{12} + ({\cal F}^3)^{\mu\nu} \Lambda^\rho
H_{\alpha\beta\rho} ({\cal F}^3)^{\alpha\beta} f_{22}.
\end{eqnarray}
This equation is analyzed in the appendix.  There it is shown that the
necessary and sufficient condition for this equation to be satisfied, given the
original field equation (\ref{fulleqn}), is that $f(y_1, y_2)$ satisfy
 the differential
equation
\begin{equation}
f_1^2 + y_1 f_1 f_2 + \left({1\over 2} y_1^2 - y_2\right) f_2^2 = 1.
 \label{diffeqn}
\end{equation}
Note that this is satisfied by the free theory $(f = y_1)$.

The differential equation can be made to look much simpler by the
change of variables
\begin{eqnarray}
y_1 &=& - (u_+ + u_-)\nonumber \\
y_2 &=& {1\over 2} (u_+^2 + u_-^2).
\end{eqnarray}
Denoting the resulting function by the same symbol, $f(u_+, u_-)$, and
derivatives by $f_\pm \equiv {\partial f\over \partial u_\pm}$, one has
\begin{eqnarray}
f_1 &=& {u_- f_+ - u_+ f_-\over u_+ - u_-}\nonumber \\
f_2 &=& {f_+ - f_-\over u_+ - u_-}.
\end{eqnarray}
Substituting these in eq. (\ref{diffeqn})  then gives the remarkably simple
differential equation
\begin{equation}
f_+ f_- = 1.
\end{equation}
Essentially the same equation was discovered in Ref. \cite{gibbons}
 as the condition for
electric-magnetic duality symmetry of a 4d $U(1)$ gauge theory.  Perhaps,
in retrospect, this
is not too surprising.

Fortunately, the general solution of the equation $f_+ f_- =1$
 is given in Courant
and Hilbert \cite{courant}.   It is given parametrically in terms of
 an arbitrary function
$v(t)$:
\begin{eqnarray}
f &=& {2u_+\over \dot v(t)} + v(t)\nonumber \\
u_- &=& {u_+\over (\dot v(t))^2} + t,
\end{eqnarray} 
where the dot means that the derivative of the function is taken
 with respect to its argument.
In principle, the second equation determines $t$ in terms of $u_+$ and $u_-$,
which can then be substituted into the first one to give $f$ in terms of $u_+$
and $u_-$.  The proof is simple, so we show it.  Taking differentials,
\begin{eqnarray}
df &=& {2\over\dot v} du_+ + \left(\dot v - {2\ddot v\over \dot v^2}u_+\right)
dt\nonumber \\
du_- &=& {1\over (\dot v)^2} du_+ + \left( 1 - {2\ddot v\over \dot
v^3}u_+\right) dt.
\end{eqnarray}
Eliminating $dt$ leaves
\begin{equation}
df = {1\over \dot v} du_+ + \dot v du_-,
\end{equation}
which implies that $f_+ = 1/\dot v$ and $f_- = \dot v$, so that $f_+ f_- = 1$.

This is not the whole story, since there is another condition that must still
be imposed.  As we have said, $f(y_1, y_2)$ is required to be analytic at the
origin.  This implies that
\begin{equation}
f (u_+, u_-) = f(u_-, u_+).
\end{equation}
So we must examine the implications of this restriction.  Since the role of
$u_+$ and $u_-$ can be interchanged in the general solution, for every $v(t)$
there must be a corresponding $w(s)$ such that
\begin{eqnarray}
f &=& {2u_-\over \dot w(s)} + w(s)\nonumber \\
u_+ &=& {u_-\over (\dot w (s))^2} + s.
\end{eqnarray}
Since $df = \dot w(s) du_+ + {1\over\dot w(s)} du_-$, we deduce that
\begin{equation}
\dot w (s) \dot v (t) = 1.
\end{equation}
Also,
\begin{equation}
u_+ = (\dot v (t))^2 (u_- - t) = {u_-\over (\dot w (s))^2} + s,
\end{equation}
then implies that
\begin{equation}
s = - t (\dot v (t))^2.
\end{equation}
Now the symmetry condition $f (u_+, u_-) = f(u_-, u_+)$ implies that $v$ and
$w$ are the same function, and therefore, $\dot v (s) \dot v (t) = 1$.  Letting
$\varphi (t) = \dot v (t)$ and substituting for $s$ then gives the functional
equation
\begin{equation}
\varphi (-t \varphi^2 (t)) \varphi (t) = 1.
\end{equation}
Letting $\psi(t) = - t \varphi^2 (t)$ (the same function as $s(t)$), the
functional equation simplifies to
\begin{equation}
\psi (\psi(t)) = t. \label{psieqn}
\end{equation}
In words, the function is the same as the inverse function.  

Large classes of solutions of (\ref{psieqn}) are obtained as follows.\footnote{We
are grateful to S. Cherkis for a discussion that helped to clarify this question.} Pick
a symmetric function $F(s,t) = F(t,s)$ and determine $\psi(t)$ by 
\[ F(\psi, t) =0. \]
For example, the simplest non-trivial choice is
\[ F(s,t) = s + t + \alpha s t ,\]
which gives
\begin{equation}
\psi (t) = {- t\over 1 + \alpha t} .
\end{equation}
One then concludes that
\begin{equation}
t = {u_- - u_+\over 1 + \alpha u_+} ,
\end{equation}
\begin{equation}
\dot v (t) = - (1 + \alpha t)^{-1/2} = - \left({1 + \alpha u_+\over 1 +
\alpha u_-}\right)^{1/2} ,
\end{equation}
\begin{equation}
v (t) = {2\over \alpha} \left[ 1 - \left({1 + \alpha u_-\over 1 + \alpha
u_+}\right)^{1/2}\right]
\end{equation}
\begin{eqnarray}
f &=& {2\over\alpha} (1 - \sqrt{(1 + \alpha u_+) (1 + \alpha u_-)}). \nonumber \\
&=& {2\over\alpha} \left(1 - \sqrt{1 - \alpha y_1 + \alpha^2 \left({1\over 2}
y_1^2 - y_2\right)} \right)\nonumber \\
&=& {2\over \alpha} (1 - \sqrt{- {\rm det} (\eta_{\mu\nu} + \sqrt{\alpha} {\cal
F}_{\mu\nu})}).
\end{eqnarray}
Reduced to 5d $({\cal F}_{\mu\nu} \rightarrow F_{\mu\nu})$ this is precisely
the Born--Infeld Lagrangian.  Henceforth we set the parameter
 $\sqrt{\alpha} = 1$.
Substituting
\begin{eqnarray}
f_1 &=& {1 - y_1\over\sqrt{1 - y_1 + {1\over 2} y_1^2 - y_2}} \nonumber \\
f_2 &=& {1\over\sqrt{1 - y_1 + {1\over 2} y_1^2 - y_2}} ,
\end{eqnarray}
the 6d field equation (\ref{fulleqn}) becomes
\begin{equation}
\tilde{H}_{\mu\nu} = {(1 - y_1) {\cal F}_{\mu\nu} + ({\cal
F}^3)_{\mu\nu}\over\sqrt{1 - y_1 + {1\over 2} y_1^2 - y_2}}.
\label{fieldequation}
\end{equation}

\section{The Lagrangian}

In the case of the free theory, we used the field equation $F_{\mu\nu} =
\tilde{H}_{\mu\nu} - \partial_5 B_{\mu\nu}$ to deduce a second order field
equation involving $B_{\mu\nu}$ only and to infer the Lagrangian that gives
this field equation.  In the case of the interacting theory, we have obtained
an equation of motion of the structure $\tilde{H}_{\mu\nu} = G_{\mu\nu} ({\cal
F})$.  In order to repeat the steps of the free theory analysis, we need to
invert this equation to one of the form ${\cal F}_{\mu\nu} = K_{\mu\nu}
(\tilde{H})$. That is what we now do.

The field equation
\begin{equation}
\tilde{H}_{\mu\nu} = {\cal F}_{\mu\nu} f_1 + ({\cal F}^3)_{\mu\nu} f_2
 \label{repeateqn} \end{equation}
can be inverted in the form
\begin{equation}
{\cal F}_{\mu\nu} = \tilde{H}_{\mu\nu} g_1 + (\tilde{H}^3)_{\mu\nu} g_2.
\end{equation}
A convenient method for making this explicit for $f = 2 (
1 - \sqrt{1 - y_1 + {1\over 2} y_1^2 - y_2})$ is to evaluate both equations in
the specific basis described in the appendix. 
 Doing this, eq. (\ref{repeateqn}) becomes
\begin{equation}
\gamma_\pm = \lambda_\pm \sqrt{{1 + u_\mp\over 1 + u_\pm}},
\end{equation}
where $u_\pm = \lambda_\pm^2$.  Also defining $h_\pm = \gamma_\pm^2$, this can
be inverted to give
\begin{equation}
\lambda_\pm = \gamma_\pm \sqrt{{1 - h_\mp\over 1 - h_\pm}}.
\end{equation}
{}From this one infers that
\begin{eqnarray}
g_1 &=& {1 - (h_+ + h_-)\over\sqrt{(1 - h_+) (1 - h_-)}} \nonumber \\
g_2 &=& - {1\over\sqrt{(1 - h_+) (1 - h_+)}}.
\end{eqnarray}
Note, in particular, that
\begin{equation}
(g_1 - h_+ g_2) (g_1 - h_- g_2) = 1. \label{useful}
\end{equation}

To recast the preceding formulas in terms of $\tilde{H}$, we define
\begin{eqnarray}
z_1 &=& {1\over 2} \tr (\tilde{H}^2) \nonumber \\
z_2 &=& {1\over 4} \tr (\tilde{H}^4),
\end{eqnarray}
and note that in the special basis $z_1 = - (h_+ + h_-)$ and $z_2 = {1\over 2}
(h_+^2 + h_-^2)$.  Substituting these formulas, one learns that $g_i =
{\partial g\over\partial z_i}, i = 1,2$, where
\begin{equation}
g (z_1, z_2) = 2 \left(\sqrt{1 + z_1 + {1\over 2} z_1^2 - z_2} - 1\right)
\end{equation}
or, equivalently,
\begin{equation}
g (\tilde{H}) = 2 (\sqrt{- {\rm det} (\eta_{\mu\nu} +
 i \tilde{H}_{\mu\nu})} - 1).
\end{equation}
We now have
\begin{equation}
F_{\mu\nu} = - \partial_5 B_{\mu\nu} + \tilde{H}_{\mu\nu} g_1 +
(\tilde{H}^3)_{\mu\nu} g_2,
\end{equation}
with $g_1$ and $g_2$ as described above.  The $F_{\mu\nu}$ Bianchi identity
then gives the desired second-order equation involving only the $B_{\mu\nu}$
field:
\begin{equation}
\epsilon^{\mu\nu\rho\lambda\sigma} \partial_\rho (\tilde{H}_{\mu\nu} g_1 +
(\tilde{H}^3)_{\mu\nu} g_2 - \partial_5 B_{\mu\nu}) = 0 .
\end{equation}
The action that gives this field equation is
\begin{equation}
S_6 = \int d^6 x ({1 \over 2} \tilde{H}^{\mu\nu} \partial_5 B_{\mu\nu}
 + g (\tilde{H})).
\label{interaction}
\end{equation}
Note that compared to the free theory action
(\ref{freeaction}), the first term is unchanged, and
the second one has a Born--Infeld-like extension.

The final thing we want to do is to demonstrate the Lorentz invariance of the
action $S_6$ directly.  In the free theory the procedure that gave the right
answer was to start with the $A_\mu = 0$ gauge formula
\begin{equation}
\delta B_{\mu\nu} = (\Lambda \cdot x) \partial_5 B_{\mu\nu} -
 x_5 (\Lambda \cdot
\partial) B_{\mu\nu}
\end{equation}
and to replace $\partial_5 B_{\mu\nu}$ by its value given by the $A_\mu = 0$
gauge field equation.  Doing the same thing again gives the formula
\begin{equation}
\delta B_{\mu\nu} = (\Lambda \cdot x) (\tilde{H}_{\mu\nu} g_1 +
(\tilde{H}^3)_{\mu\nu} g_2) - x_5 (\Lambda \cdot \partial) B_{\mu\nu}.
\end{equation}
The claim, then, is that this describes the non-manifest portion of the 6d
Lorentz invariance of $S_6$.  To check this we should show that
\begin{equation}
\delta S_6  \sim \int d^6 x \epsilon^{\mu\nu\rho\lambda\sigma}
 \delta B_{\mu\nu}
\partial_\rho (\tilde{H}_{\lambda\sigma} g_1 + (\tilde{H}^3)_{\lambda\sigma}
g_2 - \partial_5 B_{\lambda\sigma})
\end{equation}
vanishes for this choice of $\delta B_{\mu\nu}$.  In other words, we want to
demonstrate that the integrand is a total derivative.  This result has already
been demontrated for $g_1 = 1$ and $g_2 = 0$ in Section 2.

The calculation is best organized by recalling how it worked for the free
theory.  If it works the same way here, we would expect the terms linear in
$g$'s to be total derivatives and the terms quadratic in $g$'s to give a
contribution cancelling that of the term independent of $g$.  Let us begin with
the terms linear in $g$.  One of them is
\begin{eqnarray}
- \epsilon^{\mu\nu\rho\lambda\sigma} \Lambda \cdot x (\tilde{H}_{\mu\nu} g_1
+ (\tilde{H}^3)_{\mu\nu} g_2) \partial_\rho \partial_5 B_{\lambda\sigma}
&=& - 2 (\Lambda \cdot x) \partial_5 \tilde{H}^{\mu\nu} 
(\tilde{H}_{\mu\nu} g_1 +
(\tilde{H}^3)_{\mu\nu} g_2) \nonumber \\
&=& 2 (\Lambda \cdot x) (\partial_5 z_1 g_1 + \partial_5 z_2 g_2) \nonumber \\
&=& 2 (\Lambda \cdot x) \partial_5 g = {\rm tot. \, deriv.}
\end{eqnarray}
The other one is
\begin{eqnarray}
- \epsilon^{\mu\nu\rho\lambda\sigma} x_5 (\Lambda \cdot \partial) B_{\mu\nu} 
\partial_\rho (\tilde{H}_{\lambda\sigma} g_1 &+& (\tilde{H}^3)_{\lambda\sigma}
g_2)  \nonumber \\
&=& - \epsilon^{\mu\nu\rho\lambda\sigma} x_5 \Lambda^\eta H_{\eta\mu\nu}
\partial_\rho (\tilde{H}_{\lambda\sigma} g_1 + (\tilde{H}^3)_{\lambda\sigma}
g_2) + {\rm tot. \, deriv.} \nonumber \\
&=& {1\over 2} \epsilon^{\mu\nu\rho\lambda\sigma}
\epsilon_{\eta\mu\nu\alpha\beta} x_5 \Lambda^\eta \tilde{H}^{\alpha\beta}
\partial_\rho (\tilde{H}_{\lambda\sigma} g_1 + (\tilde{H}^3)_{\lambda\sigma}
g_2) + {\rm tot. \, deriv.} \nonumber \\
&=& - 2 x_5 \tilde{H}^{\alpha\beta} (\Lambda \cdot \partial)
(\tilde{H}_{\alpha\beta} g_1 + (\tilde{H}^3)_{\alpha\beta} g_2)
 + {\rm tot. \, deriv.}
\nonumber \\
&=& - 2 x_5 (\Lambda \cdot \partial) g + {\rm tot. \, deriv.} 
= {\rm tot. \, deriv.}
\end{eqnarray}

All that remains to complete the proof of Lorentz invariance of $S_6$ is to
show that the terms quadratic in $g$'s give the same contribution as in the
free theory.  The relevant expression is
\[
\epsilon^{\mu\nu\rho\lambda\sigma} \Lambda \cdot x (\tilde{H}_{\mu\nu} g_1 +
(\tilde{H}^3)_{\mu\nu} g_2) \partial_\rho (\tilde{H}_{\lambda\sigma} g_1 +
(\tilde{H}^3)_{\lambda\sigma} g_2)
\]
\begin{equation}
= - {1\over 2} \epsilon^{\mu\nu\rho\lambda\sigma} \Lambda_\rho
(\tilde{H}_{\mu\nu} g_1 + (\tilde{H}^3)_{\mu\nu} g_2)
(\tilde{H}_{\lambda\sigma} g_1 + (\tilde{H}^3)_{\lambda\sigma} g_2) +
{\rm tot. \, deriv.}
\end{equation}
The easiest way to simplify this further is to evaluate it in the special basis
described in the appendix.  Using the identity
 in eq. (\ref{useful}) one deduces that this
is the same as the free theory $(g_1 = 1, g_2 = 0)$ expression, and that,
therefore, it cancels the terms independent of $g$'s, just as in the free
theory.

\section{The String Soliton}

Since the 6d theory that has been presented here contains a
three-form field strength that is self-dual for weak fields, 
it is plausible that there is a one-brane solution
that acts as a source for both electric and magnetic charges --
{\it i.e.}, a string soliton that carries a self-dual charge.
In the free version of the theory, such a one-brane would have a
singularity in its electric field strength on the brane. Such a singularity
would lead to a configuration having infinite energy. One of the original
motivations of Born and Infeld \cite{born}
was to find theories in which such singularities were
removed, and in modern language, they showed that there is
a non-singular 0-brane in their 4d electromagnetic theory.
We will now look for an analogous one-brane in
our theory. In fact, the analogy is very close if we choose to
align the string along the $x^5$ axis and seek a solution that is independent 
of $x^5$. In this case the string soliton is mathematically the same thing as
a 0-brane soliton of the 5d Born--Infeld theory. The solution is very
similar to the 4d one of Born and Infeld.

The metric on flat 6d Minkowski space needs to split up
so that the timelike plane of the string world volume is distinguished
from the directions transverse to this plane. We therefore
 write the metric as
\begin{equation}
ds^2 = -dt^2 + (dx^5)^2 + \delta_{ab} dx^a dx^b
\end{equation}
with $a,b=1,2,3,4$. In fact the choice of cartesian co-ordinates for
the four dimensions transverse to the string is rather inconvenient
for finding solutions to the field equations. It is somewhat easier
if we rewrite the metric using a radial coordinate $\rho^2=\delta_{ab}x^ax^b$
and the line
element on the unit three-sphere $d\Omega_3$, so that we describe
Minkowski space by
\begin{equation}
ds^2 = -dt^2 + (dx^5)^2 + d\rho^2 + \rho^2 d\Omega_3^2.
\end{equation}
A convenient form for the metric on the unit three-sphere is given in
terms of Euler angles $\theta,\phi$ and $\psi$ by
\begin{equation}
d\Omega_3^2 = (d\psi+\cos\theta d\phi)^2+(d\theta^2+\sin^2\theta d\phi^2),
\end{equation}
where $0\le\theta\le\pi, 0\le\phi<2\pi, 0\le\psi<4\pi.$
Our ansatz is to set $A_0=\alpha(\rho)$ and $A_a=0$, where $\alpha(\rho)$
is some function only of $\rho$ and is to be determined. Thus, we
are talking about a string that is the source of an electric field.
It turns out that in order to be consistent, we must have some magnetic
components of the field strength tensor non-vanishing too. 
For the simple case that we are considering here, we put
$H_{\theta\phi\psi}=\beta\sin\theta$, where $\beta$ is some as yet 
undetermined constant, and all other independent components of
$H_{\mu\nu\sigma}$ are zero.  A convenient choice of $B_{\mu\nu}$ that
gives these fields is $B_{\phi\psi}=\beta(\pm 1-\cos\theta)$
with all other independent components vanishing. The ambiguity implied
by the choice of the plus or minus sign reflects the fact that our $H$
represents a source of magnetic field. If we choose the plus (minus)
sign, then the Dirac string singularity in the potential lies along the
north (south) axis running away from the three-sphere at $\rho=0$.

It is now straightforward to solve the field equation (\ref{fieldequation}).
We find that
\begin{equation}
{d\alpha\over d\rho}= {\beta\over\sqrt{\beta^2+\rho^6}}.
\label{dalphdr}
\end{equation}
This can be integrated in terms of a hypergeometric function.
Near the origin, the integrand is tending to unity so
that $\alpha \sim \rho,$ whilst as $\rho$ tends to infinity,
$\alpha \sim k+{\beta\over 2\rho^2}$ for a constant $k$.
To be precise
\begin{equation}
\alpha(\rho) = {1\over 2}\beta^{1\over 3} \biggl[{1\over 3\pi}
\Gamma({2\over 3})\Gamma({1\over 6})^2 - {3\over 5}y^{5\over 3}
{}_2F_1({5\over 6},{5\over 3};{8\over 3};y)\biggr]
\label{alpharho}
\end{equation}
where $y=(1+\rho^6\beta^{-2})^{-1}$. Comparison of this with the asymptotic
form as $\rho\rightarrow\infty$ shows that 
\begin{equation}
k = {1\over 6\pi}\beta^{1\over 3}\Gamma({2\over 3})
\Gamma({1\over 6})^2  \approx 2.226\beta^{1\over 3}.
\end{equation}
Thus we see that $\alpha(\rho)$ is
regular on the interval $0\le\rho<\infty$, and our solution is 
completely non-singular. For this reason, it deserves to be called a
soliton. 

So far, the constant $\beta$ has been completely arbitrary. However,
it defines both an electric and a magnetic charge of the string.
The magnetic charge per unit length of the string is defined to be
\begin{equation}
P=\int_{\cal C}  H,
\label{electric}
\end{equation}
where $H$ is the three-form associated with the field strength tensor
$H_{\mu\nu\rho}$ and $\cal C$ is a three sphere surrounding
the string.  
We find that
\begin{equation}
P=16\pi^2\beta.
\end{equation}
The topological nature of this charge follows from the fact that
$H=dB$ is closed.
The solution also has an electric charge per unit length of the string, 
which is determined by the field $A_0$. Equivalently, we can note that far from
the string, the system is described by the free theory with a self dual 6d $H$
field, and so
\begin{equation}
Q=\int_{\cal C} *H,
\label{magnetic}
\end{equation}
where $*$ denotes the dual of $H$. This gives
\begin{equation}
Q=16\pi^2\beta,
\end{equation}
which means that the string carries a self-dual charge.

It is well known that the Dirac-Teitelboim-Nepomechie quantization condition
\cite{dirac,teit1,nep}
restricts the charges so that
\begin{equation}
{PQ\over 2\pi}  \in {\bf  Z}.
\end{equation}
Thus the parameter $\beta$ is quantized and is given by
\begin{equation}
\beta= \pm {\sqrt{n\over 128\pi^3}}.
\end{equation}
where $n$ is a positive integer.

Lastly, we can compute the tension of the string.  Because we are 
dealing with a static solution, the action can be
identified as the energy multiplied
by the appropriate time interval. Since the (infinitely long)
string is homogeneous,
we find that the energy per unit length, which is the tension, is given by
\begin{equation}
T = 4 \pi^{3\over2}\beta^{4\over3}\Gamma({1\over 3})\Gamma({1\over 6})
\approx 332.136 \beta^{4\over3}.
\end{equation}
Thus, in contrast to the free theory, the tension is finite.

\section{Discussion}

What we have done in this paper is to construct a nonlinear generalization
of the theory of a self-dual three-form field strength. Although our analysis has
been specific to six flat Lorentzian dimensions, it
should be straightforward to extend it to spacetimes of Lorentzian
signature and dimension $4n+2$ for the case of $(2n+1)$-form field 
strengths.  For example, extended supergravity
in 9d is known from dimensional reduction of 
11d supergravity. This theory
can be lifted to Type IIB supergravity in 10d, which contains a self-dual
five-form. It should be possible to formulate a 10d action for the IIB theory,
which has manifest general covariance in 9d and a nontrivially realized
general coordinate symmetry in the tenth dimension.

The lack of an action with manifest Lorentz invariance may seem rather disturbing, 
though we are accustomed to dealing with theories having
non-manifest supersymetries. In any case,
it can be circumvented at the expense of introducing
an infinite number of auxiliary fields, as has recently been discussed
by Berkovits.\cite{berk} However, the theory described here
exhibits a new phenomenon: apparently, there is also no 
manifestly covariant form of the field equations.
We have not constructed a rigorous proof of this assertion, 
but we are reasonably confident that it is correct.  Of course,
this too might be circumvented with an infinite number of auxiliary fields.

The analysis in this paper has been entirely  classical. Recently,
Seiberg \cite{seiberg} presented evidence for the existence of
exact interacting quantum field theories in 6d.
These theories, all of which contain a chiral tensor gauge field, are inherently
non-perturbative. It seems possible that supersymmetric
extensions of our theory are somehow related to them. 
An interesting question, raised
already in the introduction, is whether in the quantum setting it makes
more sense to view the
field theory as an effective low-energy description of the self-dual string
or the self-dual string as a soliton of the more fundamental field theory.
It seems to us that Seiberg's work points toward the latter possibility.

Our primary  motivation for this work was to seek an understanding of
the M-theory five-brane. This object has amongst its world volume fields
a chiral three-form field strength. Since 
dimensional reduction must give the D four-brane, 
which contains Born--Infeld theory, we were led to the analysis
presented here. Recently, supersymmetric actions have been constructed for the
D three-brane \cite{cederwall} and for all
Type II D-branes.\cite{aganagic} This includes, in particular,
the Type IIA D four-brane,
whose world-volume theory  is an extension of 5d Born-Infeld theory. The
degrees of freedom in addition to the gauge field
are the 10d superspace coordinates. It should be possible to extend the analysis
of our paper to lift that 5d theory to a covariant 6d theory, which would
describe the M theory five-brane. Of course, general coordinate invariance would
be manifest in only five of the six dimensions.

Recently, Howe and Sezgin proposed field equations
for the M theory five-brane. \cite{howeandsezgin}
Their formalism is sufficiently different
from ours that it is very difficult to compare formulae. However,  
the fact that their equations have manifest 6d covariance in the world volume, 
contradicting our belief that this is not possible, makes us
skeptical of their results.

We wish to acknowledge the Aspen Center for Physics, where most of this work
 was done. JHS wishes to acknowledge a helpful discussion with S. Cherkis.

\vfil

\vfil
\newpage

\section*{Appendix - The Lorentz Invariance Condition}

Given the field equation
\begin{equation}
\tilde{H}_{\mu\nu} = {\cal F}_{\mu\nu} f_1 + ({\cal F}^3)_{\mu\nu} f_2,
 \label{htilde}
\end{equation}
we wish to analyze the implications of the Lorentz invariance condition
\begin{eqnarray}
{1\over 2} \epsilon^{\mu\nu\rho\lambda\sigma} {\cal F}_{\lambda\sigma} &=& -
H^{\mu\nu\rho} f_1 - H^{\mu\alpha\rho} {\cal F}_{\alpha\beta} {\cal
F}^{\beta\nu} f_2 \nonumber \\
&-& {\cal F}^{\mu\alpha} H_{\alpha\beta}{}^\rho {\cal F}^{\beta\nu} f_2 - {\cal
F}^{\mu\alpha} {\cal F}_{\alpha\beta} H^{\beta\nu\rho} f_2 \nonumber \\
&+& {\cal F}^{\mu\nu} H_{\alpha\beta}{}^\rho {\cal F}^{\alpha\beta} f_{11} +
({\cal F}^3)^{\mu\nu} H_{\alpha\beta}{}^\rho {\cal F}^{\alpha\beta} f_{12}
\nonumber \\
&+& {\cal F}^{\mu\nu} H_{\alpha\beta}{}^\rho ({\cal F}^3)^{\alpha\beta} f_{12}
+ ({\cal F}^3)^{\mu\nu} H_{\alpha\beta}{}^\rho ({\cal F}^3)^{\alpha\beta}
 f_{22}.
\label{lorentz}
\end{eqnarray}
Note that the left side has manifest $\mu\nu\rho$ antisymmetry, whereas the
right side only has manifest $\mu\nu$ antisymmetry.  Thus, there are more
conditions to satisfy than if the symmetry were manifest.

It is very convenient, and completely general, to use 5d Lorentz invariance to
map ${\cal F}_{\mu\nu}$ to a special basis in which its only nonzero components
are
\begin{eqnarray}
{\cal F}_{12} &=& - {\cal F}_{21} = \lambda_+ \nonumber \\
{\cal F}_{34} &=& - {\cal F}_{43} = \lambda_-.
\end{eqnarray}
The field equation (\ref{htilde}) then implies that $\tilde{H}_{\mu\nu}$
 has non-zero
components in the same positions, so we define,
\begin{eqnarray}
\tilde{H}_{12} &=& - \tilde{H}_{21} = \gamma_+ \nonumber \\
\tilde{H}_{34} &=& - \tilde{H}_{43} = \gamma_+ .
\end{eqnarray}
The field equation (\ref{htilde}) then gives
\begin{equation}
\gamma_\pm = \lambda_\pm (f_1 - \lambda_\pm^2 f_2), \label{gammaeqn}
\end{equation}
which can be used to eliminate $\gamma_\pm$ from the Lorentz invariance
equation.

Let us now use this special basis to study the Lorentz invariance conditions
 (\ref{lorentz}).
First consider setting $(\mu\nu\rho) = (012)$ in (\ref{lorentz}). 
 This gives the condition
\begin{equation}
(f_1 - \lambda_+^2 f_2) (f_1 - \lambda_-^2 f_2) = 1. \label{condition}
\end{equation}
However, in this special basis
\begin{eqnarray}
y_1 &=& {1\over 2} \tr {\cal F}^2 = - (\lambda_+^2 + \lambda_-^2) \nonumber \\
y_2 &=& {1\over 4} \tr {\cal F}^4 = {1\over 2} (\lambda_+^4 + \lambda_-^4) ,
\label{ydefs}
\end{eqnarray}
and thus the condition (\ref{condition}) can be rewritten in the form
\begin{equation}
f_1^2 + y_1 f_1 f_2 + ({1\over 2} y_1^2 - y_2) f_2^2 = 1. \label{thecondition}
\end{equation}
Note that this condition has $\lambda_+ \leftrightarrow \lambda_-$ symmetry,
which means that $(\mu\nu\rho) = (034)$ gives the same formula.

Because the right side of the Lorentz invariance equation 
(\ref{lorentz}) does not have total
$\mu\nu\rho$ antisymmetry manifest, $(\mu\nu\rho) = (120)$ must be analyzed
separately.  It gives
\begin{eqnarray}
\lambda_- &=& \gamma_- f_1 - 3\gamma_- \lambda_+^2 f_2 - 2\lambda_+ (\gamma_-
\lambda_+ + \gamma_+ \lambda_-) (f_{11} - \lambda_+^2 f_{12})\nonumber \\
&&  + 2 \lambda_- (\gamma_- \lambda_+^3 + \gamma_+ \lambda_-^3) (f_{12} -
\lambda_+^2 f_{22}).
\end{eqnarray}
Eliminating $\gamma_\pm$ using eq. (\ref{gammaeqn}) leaves
\begin{eqnarray}
0 &=& (f_1 - \lambda_-^2 f_2) (f_2 + f_{11} - 2\lambda_+^2 f_{12} + \lambda_+^4
f_{22}) \nonumber \\
&& + (f_1 - \lambda_+^2 f_2) (f_{11} - (\lambda_+^2 + \lambda_-^2) f_{12} +
\lambda_+^2 \lambda_-^2 f_{22}).
\end{eqnarray}
The $\mu\nu\rho = (340)$ equation is the same with $\lambda_+ \leftrightarrow
\lambda_-$.  It is convenient to form the sum and difference of the two
equations.  The difference equation is
\begin{equation}
0 = - 2 f_1 f_{12} + f_2^2 + (\lambda_+^2 + \lambda_-^2) (f_1 f_{22} + f_2
f_{12}) - 2 \lambda_+^2 \lambda_-^2 f_2 f_{22}
\end{equation}
or using (\ref{ydefs})
\begin{equation}
2f_1 f_{12} + y_1 (f_{12} f_2 + f_1 f_{22}) - f_2^2 + (y_1^2 - 2 y_2) f_2
f_{22} = 0.
\end{equation}
Remarkably, this is just what one obtains from differentiating
 eq. (\ref{thecondition})
with respect to $y_2$.  So no additional constraint arises.  The sum of the two
equations gives
\begin{eqnarray}
&&2 f_1 f_2 + 4 f_1 f_{11} + y_1 (f_2^2 + 2 f_{11} f_{22} + 4 f_1
f_{12})+ 2 y_2 (f_1 f_{22} + f_2 f_{12}) \nonumber \\
&& + (y_1^2 - 2 y_2) (3f_2 f_{12} + f_1
f_{22}) +  y_1 (y_1^2 - 2 y_2) f_2 f_{22} = 0.
\end{eqnarray}
This equation is automatically satisfied using both the $y_1$ and $y_2$
derivatives of eq. (\ref{thecondition}).  Hence eq. (\ref{thecondition})
 is all that is required.

\vfil

\end{document}